\documentclass[iop]{emulateapj}

\usepackage{epsfig}
\usepackage{graphicx}
\newcommand{\kms}{\ifmmode {\rm km\,s}^{-1} \else km\,s$^{-1}$\fi}

\shorttitle{CH$_{3}$OH Maser Conditions}
\shortauthors{McEwen et al.}

\begin{document}

\title{Class I Methanol (CH$_{3}$OH) Maser Conditions near Supernova
  Remnants}

\author{Bridget C. McEwen, Ylva M. Pihlstr\"om\footnote{Y.\ M.\
    Pihlstr\"om is also an Adjunct Astronomer at the National Radio
    Astronomy Observatory}}
\affil{The Department of Physics and Astronomy, The University of New Mexico, Albuquerque, NM, 87131}

\author{Lor\'{a}nt O. Sjouwerman}
\affil{National Radio Astronomy Observatory, P.O. Box O, 1003 Lopezville Rd., Socorro, NM, 87801}

\begin{abstract}
  We present results from calculations of the physical conditions
  necessary for the occurrence of 36.169 ($4_{-1}-3_{0}\, E$), 44.070
  ($7_{0}-6_{1}\,A^+$), 84.521 ($5_{-1}-4_{0}\,E$), and 95.169
  ($8_{0}-7_{1}\,A^+$) GHz methanol (CH$_3$OH) maser emission lines
  near supernova remnants (SNRs), using the MOLPOP-CEP program. The
  calculations show that given a sufficient methanol abundance,
  methanol maser emission arises over a wide range of densities and
  temperatures, with optimal conditions at $n\sim 10^4-10^6$ cm$^{-3}$
  and $T>60$ K. The 36~GHz and 44~GHz transitions display more
  significant maser optical depths compared to the 84~GHz and 95~GHz
  transitions over the majority of physical conditions. It is also
  shown that line ratios are an important and applicable probe of the
  gas conditions. The line ratio changes are largely a result of the
  $E$-type transitions becoming quenched faster at increasing
  densities. The modeling results are discussed using recent
  observations of CH$_3$OH and hydroxyl (OH) masers near
  the supernova remnants G1.4$-$0.1, W28, and Sgr\,A\,East.

\end{abstract}

\section{Introduction}

Studying maser lines is useful in order to explore the physical and
chemical conditions in interstellar gas. Maser pumping occurs under
specific conditions, and modeling of the maser formation can provide
estimates of the required environmental parameters. If more than one
transition can be observed at the same location, predictions of line
ratios can be used to place tight constraints on the physical
conditions. A suitable species for studies is therefore the methanol
(CH$_3$OH) molecule, which displays several maser transitions that can
be found in the interstellar medium \citep{cragg1992}. Methanol masers
were first discovered by \citet{barrett1971} at
frequencies around 25~GHz near Orion. Since then, methanol maser
emission has been discovered in hundreds of star forming regions
\citep[e.g.,][]{bachiller1990, beuther2002, kurtz2004}. As originally
suggested by \citet{batrla1987}, and later labelled by
\citet{menten1991}, methanol masers are divided into two classes which
depend on their pumping mechanism. Class I methanol masers are
collisionally pumped, whereas the Class II masers are pumped by
radiation \citep[e.g.,][]{cragg1992,cragg2002}. Both classes can be
found associated with massive star forming regions (SFRs). Consistent
with the pumping classification, Class II methanol masers are typically
located close to the protostars, often associated with UC and HC HII
regions, while the Class I masers instead are related to outflows and
molecular shock fronts (\citealt{plambeck1990, kurtz2004,
  voronkov2006}).

Masers are not only excited in SFRs but also in regions shocked by
supernova remnants (SNRs). For instance, the collisionally pumped
1720~MHz~OH masers are associated with the interaction regions between
SNRs and molecular clouds (e.g., \citealt{claussen1997, frail1998,
  yusef2003}). A comparison of the OH and methanol molecule excitation
mechanism indicates that conditions conducive for 1720 MHz OH masers
may also be favorable for Class I methanol masers
\citep{lockett1999,pihl2008,pratap2008}. Based on these results it
would not be unreasonable to expect methanol masers to occur near SNRs
that sustain OH masers. Consistent with this prediction, 36~GHz and
44~GHz methanol masers have been found near the SNR Sgr\,A\,East in
the Galactic center region
(\citealt{yusef2008,sjou2010,pihl2011}). Additionally, in a recent
study by \citet{pihl2014}, the same methanol transitions were
discovered toward the SNRs G1.4$-$0.1 and W28. These three SNRs all
harbor OH masers, and therefore readily provide a testbed for combined
modeling with observations of several maser species to determine local
densities and density gradients. Constraining density estimates in
these interaction regions may be of interest for, e.g., modeling of
cosmic ray acceleration in SNRs. Such models estimate the expected
resulting $\gamma$-ray emission due to neutral pion decay, which is a
function of the number density
\citep[e.g.,][]{drury1994,abdo2010,cristofari2013}.

In this paper, modeling of the methanol maser transitions is used to
predict under what physical conditions masers will occur, specific to
a SNR environment. These results are used to interpret
recent observational results on methanol masers near SNRs.

\section{Model Parameters}

Given that methanol typically has been searched for and found associated with hot dense
cores of newly forming massive stars \citep{muller2004}, most
theoretical work on the methanol molecule and its excitation has been
focused on conditions in SFRs and Class~II masers. For the Class~I
masers, an early statistical equilibrium calculation for the 36~GHz
and 44~GHz lines detected in Sgr\,B2 showed that high kinetic
temperatures were required for maser action \citep{mori1985}. Later,
more detailed modeling of the general behaviour of both Class~I and II
masers were performed by \citet{cragg1992}. Their work concluded that
the brightest collisionally pumped Class~I methanol masers are
expected at the transitions at 36.169~GHz ($4_{-1}-3_{0} E$) and
44.070~GHz ($7_{0}-6_{1} A^+$), with slightly weaker masers for the
84.521~GHz ($5_{-1}-4_{0} E$) and 95.169~GHz ($8_{0}-7_{1} A^+$)
transitions. These four transitions have been observed associated with
outflows in SFRs (e.g., \citealt{kurtz2004, fish2011, fontani2010,
  kalenskii2013}). In SNRs, the 84~GHz line has not yet been reported,
but the 36~GHz, 44~GHz, and 95~GHz lines have all been detected
\citep{yusef2008,sjou2010,pihl2011,shulga2012}. The
occurrence of these lines, and the relative line intensities for
spatially coincident emission allows the physical conditions of the gas to be
probed.

In this paper, we examine the conditions regulating maser activity
near SNRs. This environment includes radiation due to the cosmic
microwave background (CMB), and a dust generated radiation field that
is weaker than in SFRs. The optical depths of the four Class I
transitions were calculated using the MOLPOP-CEP\footnote{Available at
  http://www.iac.es/proyecto/magnetism/pages /codes/molpop-cep.php}
program \citep{elitzur2006}. MOLPOP-CEP is a Fortran 90 code that
solves the molecular level population equations along with the
radiative transfer equations for multi-level systems.  It applies
either the Coupled Escape Probability (CEP) method outlined in
\citet{elitzur2006} or the standard escape probability approximation
for a homogeneous slab.

In order to calculate the masing conditions in MOLPOP-CEP, several
variables can be adjusted. These include the number of energy levels
considered, the kinetic gas temperature ($T$), the molecular hydrogen
density ($n$), the fractional abundance of a trace molecule ($f$, here methanol), and
radiation fields due to dust and the CMB. SNR dust temperatures are
estimated to be approximately between 20-80~K
\citep[e.g.,][]{andersen2011,lee2011}, but a change within this modest
range of dust temperatures does not significantly affect our model results. An
external radiation field of 2.725~K from the CMB and a 30~K dust radiation
field is therefore applied in each model presented here. Energy levels
for the methanol molecule were incorporated using the Leiden Atomic
and Molecular Database (LAMDA)\footnote{Available at
  http://home.strw.leidenuniv.nl/$\sim$moldata/}, which includes close
to 300 rotational energy levels for each of the A- and E-type species in the torsional ground state up to about 350 cm$^{-1}$ ($j_{max}=20$) \citep{schoeier2011}. The collision rate coefficients
used for methanol colliding with H$_2$ were also adopted from LAMDA,
originally calculated by \citet{pottage2004} for temperatures ranging
from 5 to 200~K. Since methanol masers are predicted to be excited
under similar, though not identical, conditions as the OH masers, the
initial input parameters to the models were those optimally producing
1720 MHz OH maser emission in SNRs: $n=10^5$~cm$^{-3}$ and $T=75$~K
\citep{lockett1999, pihl2008}.

\begin{deluxetable}{ccrr}
\tabletypesize{\scriptsize}
\tablecaption{Parameter Ranges for  Each Transition which causes $\tau<0$ \label{tbl-1}}
\tablewidth{0pt}
\tablehead{
\colhead{Transition} & \colhead{n Range} & \colhead{T Range}& \colhead{Max $\tau$} \\ 
\colhead{(GHz)} & \colhead{(cm$^{-3}$)}  & \colhead{(K)} & \colhead{(inverted)}}
\startdata
36 & $10^2 - 10^7$ & 20       & 3.3 \\
   & $10^2 - 10^8$ & 40 - 100 & 6.3\\
   & $10^2 - 10^9$ & 120 - 200& 6.8\\
\hline
44 & $10^5 - 10^7$ & 40       & 4.5\\
   & $10^3 - 10^7$ & 60       & 5.0\\
   & $10^2 - 10^7$ & 80       & 5.3\\
   & $10^2 - 10^8$ & 100 - 180& 5.8\\
   & $10^3 - 10^8$ & 200      & 5.9\\
\hline 
84 & $10^3 - 10^6$ & 20       & 0.6\\
   & $10^3 - 10^7$ & 40       & 1.8\\
   & $10^2 - 10^8$ & 60 - 180 & 4.2\\
   & $10^2 - 10^9$ & 200      & 4.2\\
\hline
95 & $10^5 - 10^7$ & 40       & 2.0\\
   & $10^3 - 10^7$ & 60       & 2.5\\
   & $10^2 - 10^7$ & 80 - 100 & 3.0\\
   & $10^2 - 10^8$ & 120 - 180& 3.5\\
   & $10^3 - 10^8$ & 200      & 3.6\\
\enddata
\end{deluxetable}

\section{Results}\label{results}

\subsection{Modeling Results}

To determine the full range of conditions under which masing occurs
for each transition, different combinations of $T$, $n$, and $f$ were
modeled.  The models covered a temperature range of 20 to 200~K, a
density range of 10 to 10$^9$ cm$^{-3}$, and a fraction range of
10$^{-5}$ to 10$^{-8}$. Even though $f$ was varied in the
calculations, this value only affects the absolute value of the
optical depth, and does not alter the range of temperatures and
densities under which masers form. A value of $f=10^{-7}$ is therefore
used in the remainder of this paper. All parameters were varied
successively by factors of 10, covering an extensive combination of
parameters. The number of transitions involved in the calculations can
also be varied, but this has no significant effect on our modeling outcome since
the conditions modeled here predominantly affect the lower energy
transitions only.

For OH maser excitation, line overlap is an important factor
\citep[e.g.,][]{gray1991,pavlakis1996,gray2001}, which may drastically
change the relative excitation population of the OH main line masers
\citep{cragg2002}. Line overlap in methanol is expected to be much
less important, as has been previously confirmed through modeling by
\citet{rollig1999} and \citet{cragg2002}. Due to the dominant collisional excitation mechanism in SNRs, only small changes in the resulting calculated emission would be expected due to IR line overlap given the weak dust radiation field of 30K. In fact, using line overlap for a linewidth of 1~km\,s$^{-1}$ causes a slight shift of the peak optical depth to
higher column densities by a factor of $5\%$ or less.  No other major
differences in our modeling output are found, in agreement with the
earlier modeling of methanol masers in SFRs. In the description of the
results below, line overlaps are excluded.

Varying the main parameters $T$ and $n$ results in a wide set of
optical depths, summarized in Table 1. The values in Table 1
emphasize the broad range of conditions under which the methanol
maser may be formed. The lower energy level transitions at 36~GHz and
44~GHz typically show more significant optical depths compared to
84~GHz and 95~GHz. An example of this general behavior can been seen
in Fig.\,1 which shows the inverted optical depths (i.e.,
$-1\times\tau$) at a specific set
of conditions ($n=10^5$~cm$^{-3}$ and $T=80$~K).

\begin{figure} 
\includegraphics[width=\columnwidth]{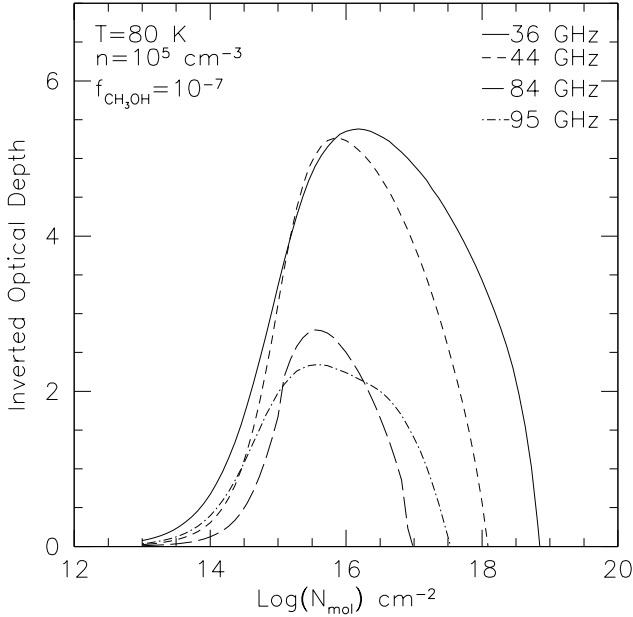}
\caption{Inverted optical depth versus the column density of
  the four methanol transitions. The lower energy level transitions at
  36~GHz and 44~GHz typically show more significant optical depths
  compared to 84~GHz and 95~GHz.}
\end{figure}

Figure 2 demonstrates how the peak inverted optical depth increases
with density between $10^2-10^4$\,cm$^{-3}$, maximizes at
$10^5-10^6$\,cm$^{-3}$, and lastly decreases drastically above
$10^7$\,cm$^{-3}$. From the same plot it can also be noted that the
$A^+$-type 44~GHz and 95~GHz transitions have the highest inverted
optical depth at a density of $10^5$\,cm$^{-3}$, while this occurs at a
slightly higher density of $10^6$\,cm$^{-3}$ for the $E$-type 36~GHz
and 84~GHz lines. Consequently, the $A^+$-type transitions get
quenched at a smaller density than the $E$-type transitions.

These trends occur for each value of the temperature, and is clearly
visible in Fig.\,3 showing contour plots of the optical depth as a
function of the H$_2$ density and the methanol column density at two
different temperatures. The 36~GHz and 84~GHz lines show optimal maser
output at higher densities and at a slightly larger column density
compared to the 44~GHz and 95~GHz lines respectively. This figure also
shows that at higher temperatures, the peak densities approach each
other, indicated by the cross and plus signs.

\begin{figure} [h!]
\includegraphics[angle=0,scale=.43]{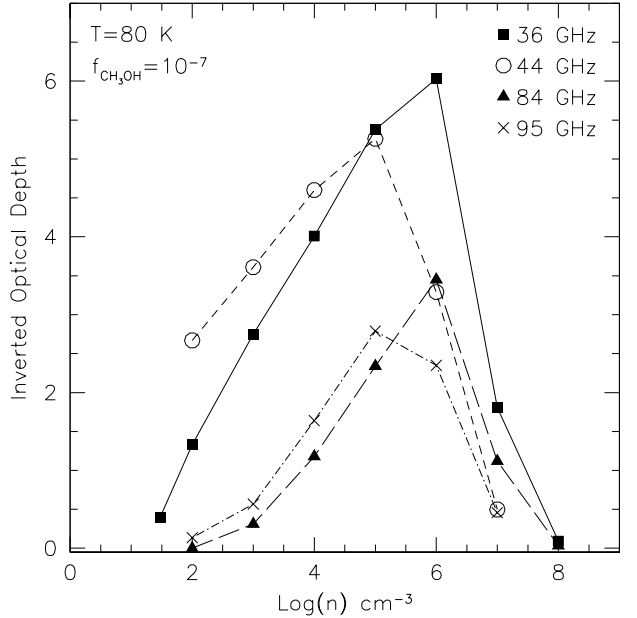}
\caption{Peak inverted optical depth versus number density of H$_2$.
  The filled symbols represent $A^+$-type transitions and the open
  symbols represent $E$-type transitions. The $E$-type transitions
  have more significant optical depths compared to the $A^+$-type
  transitions at lower densities, and conversely at higher densities.}
\end{figure}

\begin{figure*}
\center
\includegraphics[angle=0,scale=.83]{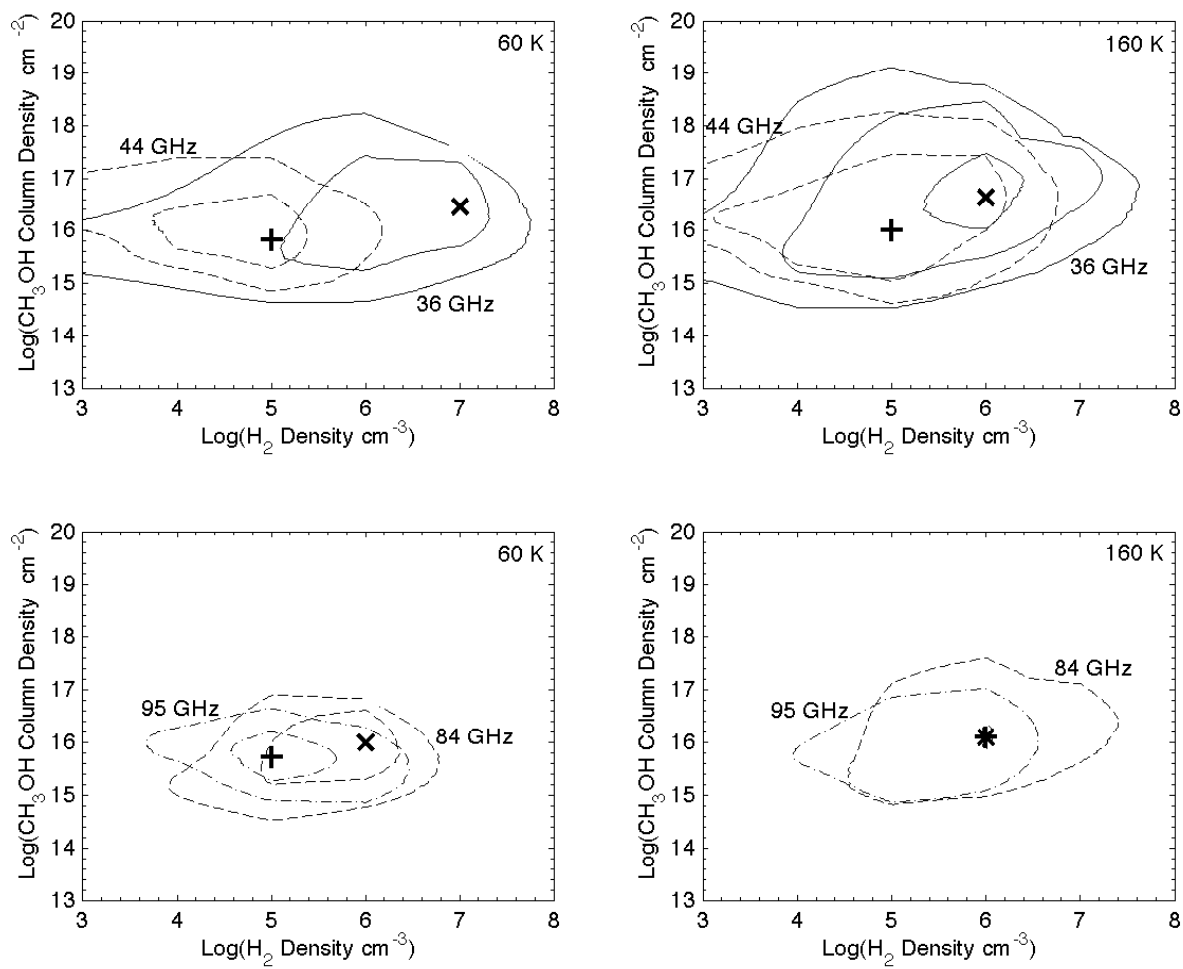}
\caption{Contour diagrams of the maser optical depth as a function of
  H$_2$ density and methanol column density. The contour levels
  correspond to peak inverted optical depths of 2, 4 and 6. The
  crosses mark the position of the highest peak inverted optical
  depth for the $A^+-$type transitions. The pluses mark the position of the highest peak inverted optical
  depth for the $E-$type transitions. Optimal masing H$_2$ densities for the $A^+-$type transitions
  (44~GHz and 95~GHz) are lower than those for the $E$-type
  transitions (36~GHz and 84~GHz).}
\end{figure*}

The temperature dependence is illustrated in Fig.\,4, using three
different densities. Overall, for a given density the optical depth
increases with increasing temperature. However, there are
parameter regions where the relative strength of the lines changes,
which is an important observational diagnostic of the region. In
particular, the $A^+$-type transitions become quenched at lower
densities than the $E$-type transitions, leaving strong masers at
36~GHz and 85~GHz at densities $\geq 10^7$\,cm$^{-3}$. At lower
densities the 44~GHz line will dominate. For more detailed diagnostic
purposes, illustrating this behavior, the optical depth ratios with respect to the 44~GHz and 36~GHz lines at
three different temperatures are plotted in Fig.\,5.

\begin{figure*}[tbh]
\includegraphics[angle=0,scale=.86]{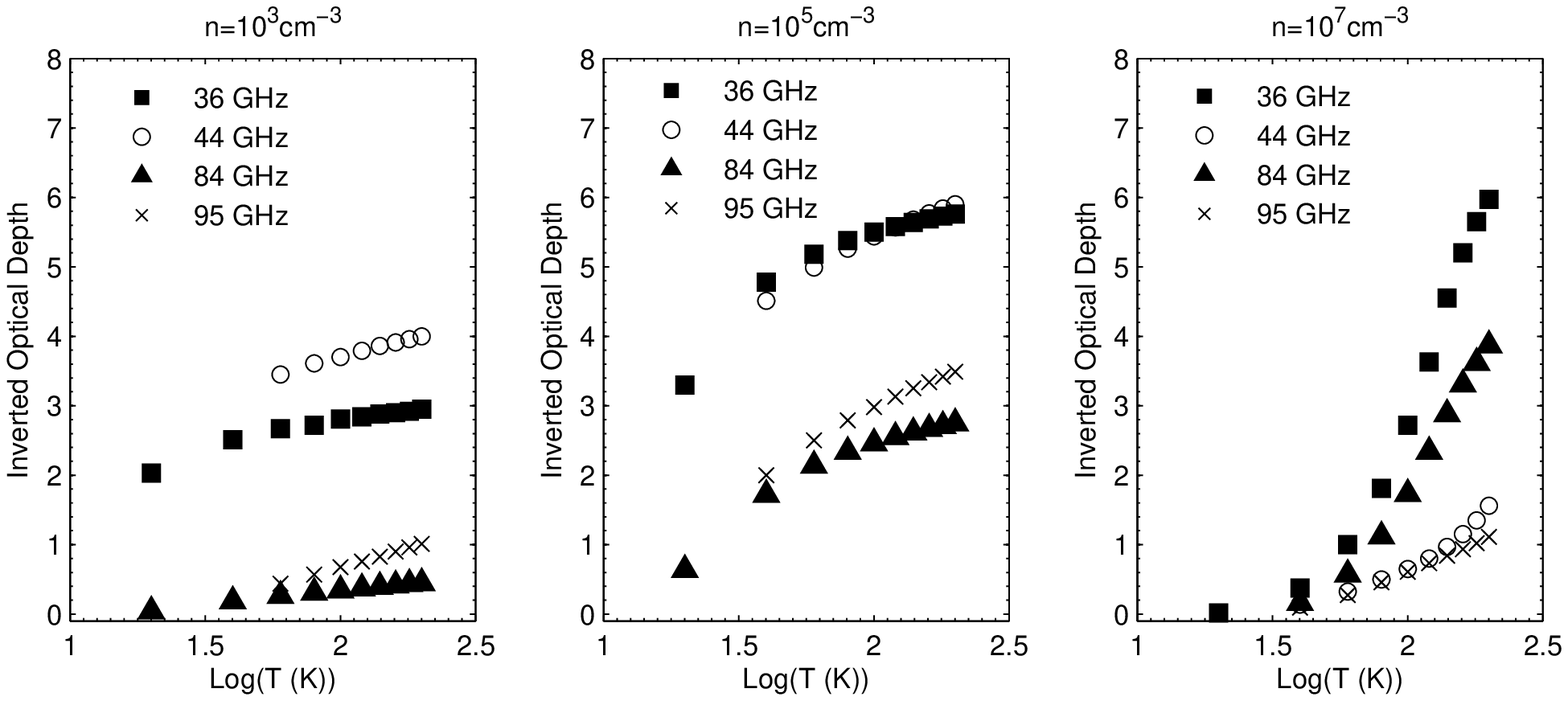}
\caption{Peak inverted optical depth versus temperature at three
  different densities. An inversion of the line ratio for the 36~GHz
  and 44~GHz may occur when the temperature increases (middle
  panel). At higher densities (right hand panel) the 44~GHz
  transitions starts to become quenched, and instead the 84~GHz line
  becomes more prominent. Note that for the lower temperatures
  ($\log(T)$ below about 1.6 or 1.7), the $A^+-$type transitions have
  positive opacities (i.e., negative inverted opacities) and are thus 
  not masing.}
\end{figure*}

\section{Discussion}

Typically, for $n < 10^7$ cm$^{-3}$, the brightest masers occur at
36~GHz and 44~GHz and therefore are most likely the most useful
observational diagnostic for probing physical conditions in the
observed region.

\subsection{Comparison between 36~GHz and 44~GHz}

Optimal masing conditions at 36~GHz occurs at temperatures $\geq 50$ K
and at densities between $10^5-10^7$ cm$^{-3}$, while optimal
conditions for the 44~GHz line are found at temperatures $\geq 50$~K
and at slightly lower densities between $10^4-10^6$ cm$^{-3}$. The
large overlap in the density range for the two transitions suggests
that 36~GHz and 44~GHz masers can be expected to be found
co-spatially \citep{pihl2011}. If only 36~GHz masers are detected, the density is
most likely too high for the formation of 44~GHz masers. If the two
transitions are found jointly, their relative brightness may be used
to constrain the density and temperature at that position (see
Sect.\,\ref{gc}).

\subsection{Comparison to Methanol in SFRs}
\citet{pratap2008} mention the results of unpublished methanol maser
modeling in SFRs suggesting that the 36 GHz transitions occur for gas
temperatures that range from about 30$-$100~K and densities on the
order of $10^{4}-10^{5}$ cm$^{-3}$. The same unpublished modeling
results reported for the 44~GHz transitions are found to occur at
higher temperatures (80$-$200 K) and densities ($10^{5}-10^{6}$
cm$^{-3}$). They predict that in SFRs the density trend is opposite to
that of our modeling results in SNRs, as we find that 44~GHz masers tend to
occur at slightly lower densities than the 36~GHz masers.
We note, however, that the unpublished \citet{pratap2008} results
should be verified before being used in a detailed quantitative
analysis.

Although 84~GHz and 95~GHz methanol masers have been detected in SFRs
(e.g., \citealt{plambeck1990, kalenskii2013}), there are less details
available for estimates of the density and temperature for these
transitions \citep{cragg1992}. The results presented here show that
these transitions may be found under the same range of
conditions as the 36~GHz and 44~GHz lines. These higher 
frequency masers are generally less bright, but they still provide
complimentary information about the interstellar conditions.

\subsection{36~GHz and 44~GHz Methanol Masers in SNRs}
The broad range of conditions under which methanol masers can form
encompasses those of the 1720 MHz OH maser line. This suggests that
the methanol lines may be readily detectable near SNRs, and perhaps
especially near those harboring OH masers. In a targeted search 
towards a sample of 21 SNRs displaying the 1720 MHz
OH maser line, which may indicate a molecular-cloud interaction, 
three SNRs (G1.4$-0.1$, W28, and Sgr\,A\,East) were
found to have 36~GHz and/or 44~GHz methanol masers 
\citep{pihl2014}. While the low number
of sources detected in any of the methanol transitions were somewhat
surprising, \citet{pihl2014} discuss some possible explanations. The
most selective one is the small angular area covered in the search, due
to the small field-of-view at those high frequencies. Also the limited
range in line-of-sight velocity may have played a role. Alternatively,
the methanol abundance or path-length may simply be insufficient for maser
amplification. Finally, if the conditions across the pre-shock to
post-shock regions changes significantly, the formation and
destruction mechanisms of the different molecules could result in a
significant spatial/angular and velocity offset between OH and methanol, and
the search was targeting the OH maser positions
specifically. Nevertheless, for the three SNRs with methanol maser
emission, the nature of the masers and the gas conditions are
discussed using the modeling results (Sect.\,\ref{g1.4}-\ref{gc}).

\subsubsection{G1.4-0.1}\label{g1.4}
Close to 40 methanol masers were detected in
G1.4$-$0.1 at 36 GHz. This SNR is one of the smaller angular scale SNRs observed by
\citet{pihl2014}, which may have increased the chance of detection.
However, most pointings contained several maser spots, indicating
there may be more masers outside the pointing areas and that this SNR
may be particularly rich in methanol masers. Consistent with this,
G1.4$-$0.1 is overlapping with two molecular clouds along the line of
sight, as probed by both CO and CS observations showing emission
concentrated around $+30$ and $-30$ \kms\, for the two clouds
respectively \citep{oka1998,tsuboi1999}. Given the close agreement of
the CO, CS, and 36~GHz methanol emission velocities it is therefore
reasonable to assume the methanol maser emission is associated with
gas in these clouds, presumably pumped by the interaction between the
SNR and the molecular clouds. The velocity of the G1.4$-$0.1 OH maser
is offset from the two main methanol velocities, perhaps because it
arises in a region of the gas where the velocities have been disturbed
on the post-shocked region \citep{pihl2011,pihl2014}. Using the MOLPOP-CEP
results (Sect.\,\ref{results}), regions where 36~GHz masers are much brighter
than the 44~GHz transition correspond to high density and
temperatures, $n\sim 10^{6}-10^{7}$\,cm$^{-3}$ and $T\geq 60$\,K
(Fig.\,4). The non detection of the 44~GHz masers can be used to
further constrain the density and temperature, using the 36~GHz maser
flux densities and rms sensitivity quoted in \citet{pihl2014}. The
average flux density of the 36~GHz masers is 0.21 Jy\,beam$^{-1}$, and
with a 10~mJy rms sensitivity a 3-$\sigma$ detection limit on the
44~GHz masers would correspond to a minimum optical depth ratio of
7. From Fig.\,5, this best corresponds to the peak region at $n\approx
10^7$\,cm$^{-3}$ and $T>140$\,K. The presence of high density regimes in
this source is confirmed by rich emission from both SiO
\citep{riquelme2010} and NH$_3$. Likewise,
Fig.\,5 can be used to predict masers in the 84~GHz transition of
$\approx 13$\,mJy\,beam$^{-1}$. Confirming this observationally would be
a good test of the model calculations.

\begin{figure*}[thb]
\includegraphics[angle=0,scale=0.44]{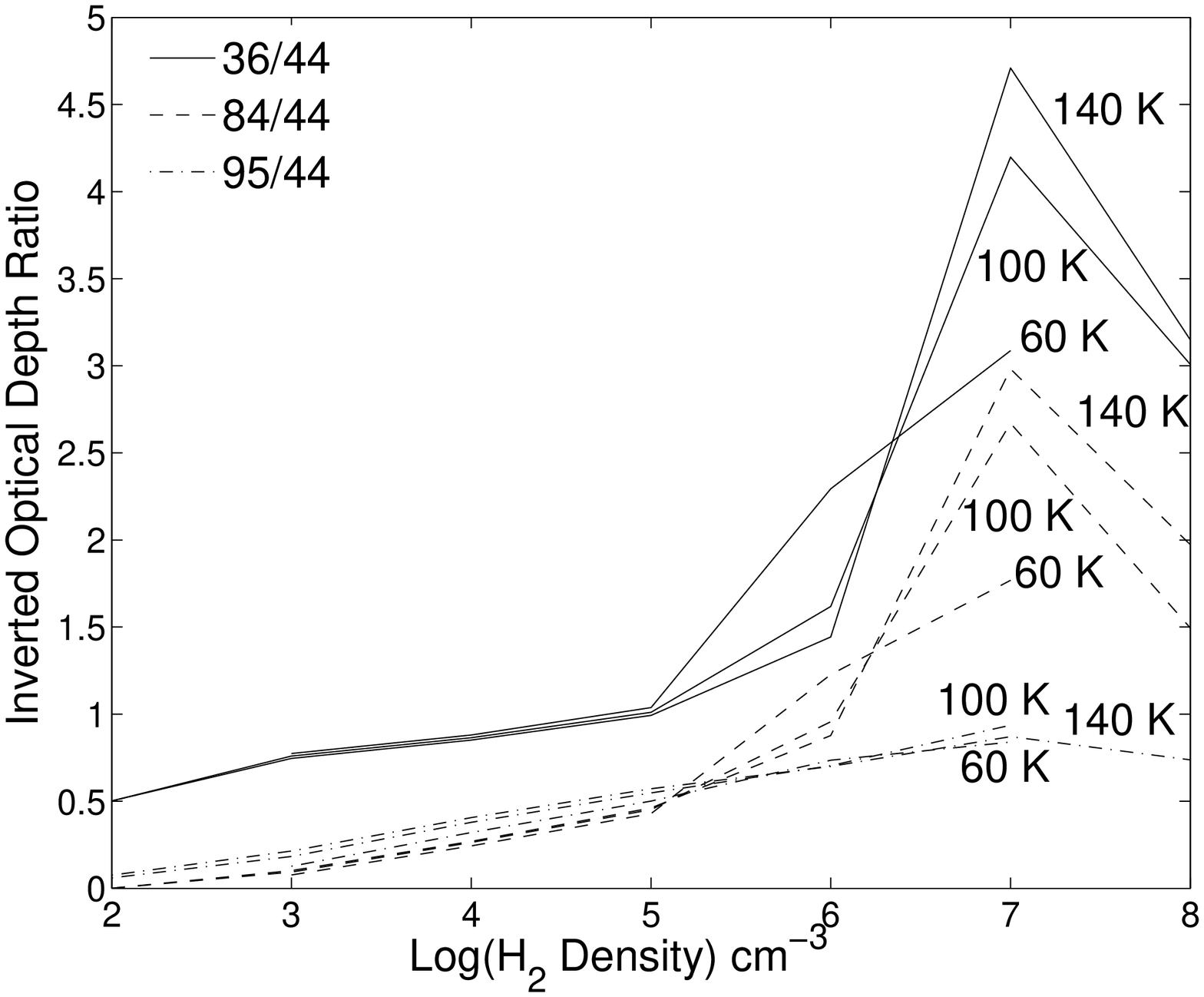}
\includegraphics[angle=0,scale=0.44]{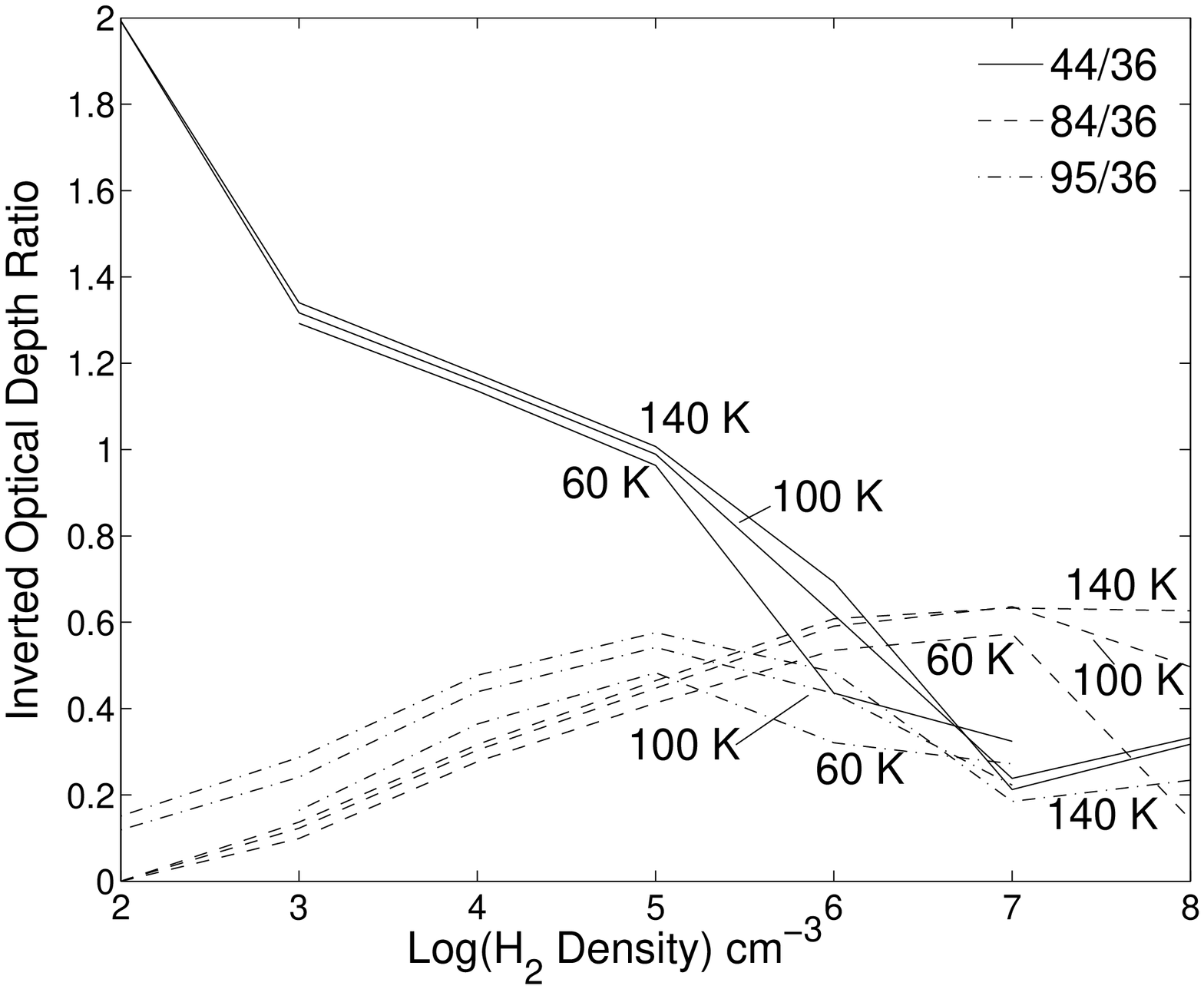}
\caption{Optical depth ratio as a function of H$_2$ density at three
  different temperatures, with respect to the 44~GHz transition (left
  panel) and 36~GHz transition (right panel). Only the 44~GHz maser
  can be stronger than the 36~GHz maser, and only for densities below
  about $10^5$ cm$^{-3}$.  The lines are drawn for the masing conditions listed in Table 1.}
\end{figure*}

\subsubsection{W28}\label{w28}
W28 and G1.4-0.1 are similar in linear sizes (27.9 and 24.7 pc respectively), but W28 is closer (2 kpc compared to 8.5 kpc for G1.4-0.1) and therefore W28 has a substantially larger angular size ($45'\times45'$ versus $10'\times10'$).  Consequently, a smaller angular area of W28 (2$\%$) was searched for 36 and 44~GHz methanol masers
\citep{pihl2014}. Two 36~GHz and four 44~GHz masers were found, in
contrast to the wealth of 36~GHz methanol emission seen in G1.4$-$0.1. The
44~GHz and 36~GHz masers are not co-spatial but instead observed in two
distinct groups. Each group is offset from, but nearby, known OH
masers, consistent with the methanol masers being associated with a
shocked environment.

The clumping of the two transitions implies one lower- and one
higher-density region. First, the 44~GHz maser region corresponds to a
position of bright NH$_3$ (3,3) inversion line emission
\citep{nicholas2011} with a critical density close to 10$^4$
cm$^{-3}$. This limits the density regime to $10^4-10^5$ cm$^{-3}$,
since at higher densities 36~GHz masers would be expected with similar
brightness as the 44~GHz masers. For the clump of 36~GHz emission, the
density may be higher ($n>10^6$ cm$^{-3}$), suppressing the 44~GHz
transition. Both of these regions are associated via shocks as shown
by the presence of OH masers as well as SiO emission
\citep{goss1968,frail1994,nicholas2011}, and the grouping of methanol
masers implies the presence of clumps and/or density gradients.

\subsubsection{The Galactic Center Region}\label{gc}
The Galactic Center (GC) is a complex region where at least two SNRs
are present (Sgr\,A\,East and G359.02$-$0.09). The Sgr\,A\,East SNR is
interacting with the two main molecular clouds, M$-$0.02$-$0.07
(a.k.a.\,the 50 km\,s$^{-1}$ cloud) and M$-$0.13$-$0.08 (a.k.a.\,the
20 km\,s$^{-1}$ cloud), providing collisionally compressed regions. In
this rich environment, four different collisionally excited maser
types have been found: 22~GHz H$_2$O, 1720 MHz OH, 36~GHz CH$_3$OH,
and 44~GHz CH$_3$OH
\citep{yusef_1995,yusef2008,sjou2010,pihl2011,pihl2011_b}. Figure 6
outlines the distribution of methanol and OH masers in the GC as
published originally by \citet{pihl2011}, using five selected pointing
positions. Three regions of methanol masers may be identified, to be
discussed in turn: the southern part with a couple of
36~GHz masers (a), a line of co-spatial 36~GHz and 44~GHz masers in
the north-east (b), and finally a region in the north-east corner
dominated by 44~GHz emission (c).

Region a) in the south, which
as in the case of G1.4$-$0.1 (Sect.\,\ref{g1.4}), can be interpreted as
a high density, high temperature region with $n\sim 10^7$ cm$^{-3}$
and $T\geq 100$ K (Figs.\,4 and 5).

Region b) is identified in \citet{pihl2011} as a sharp line following
the edge of the radio continuum contour found delineating the SNR shock
front. Here, the 36~GHz masers are always much brighter than the 44~GHz masers,
with flux density ratios between 10-75. Based on our modeling, this
implies that the densities of the region are relatively high, close to
$10^7$ cm$^{-3}$, and thus closely corresponding to values found in
region a) as well as in G1.4$-$0.1 (Sect.\,\ref{g1.4}).

Region c) differs somewhat from the previous cases but resembles
the group of 44~GHz masers found in W28 (Sect.\,\ref{w28}). One
possibility is that these masers originate closer to the inner regions
of the molecular cloud, not yet disturbed by the shock with
densities $n<10^5$ cm$^{-3}$. In this case, they may be associated
with star formation. However, an alternative scenario is offered by
the recent discovery of a dense molecular shell in this region,
outlined by observations in the CS (J=1-0) line
\citep{tsuboi2012}. This shell is also identified via a high
SiO/H$^{13}$CO$^+$ ratio \citep{tsuboi2011}, indicating a shocked
shell. \citet{tsuboi2011} suggest that this is a young ($\sim 4\time10^4$ years) SNR expanding at a velocity of about 25\,km\,s$^{-1}$. Such a shock velocity is
consistent with a C-shock, allowing the formation of methanol
molecules. To detect CS, the density has to exceed the critical density of around $10^4$ cm$^{-3}$. The
positions of the 44~GHz methanol masers follow the dense ridge of this
molecular shell \citep{sjou2010,pihl2011}, suggesting that these masers
may be excited collisionally by a different SNR shock front than those
in a) and b). This scenario is currently being investigated in more
detail by mapping a more complete region of the GC in both 36 and 44~GHz methanol
transitions.

Both methanol transitions in the high-density regions a) and b) are
offset spatially (and in velocity) from the 1720~MHz OH, which can be understood by the
OH maser being most effective at more intermediate densities of
$n\sim10^4-10^5$\,cm$^{-3}$. However, the optical depth calculations
presented in Figs.\,4 and 5 allow equally bright 36~GHz and 44~GHz
methanol at these intermediate densities, which is not observed. The
lack of 36~GHz and 44~GHz masers agrees with the notion that OH
molecules are being formed downstream the shock wave, and consequently
OH masers are found in the post-shock region where methanol molecules
may have been destroyed. The passage of a shock can efficiently enable
O atoms for the production of H$_2$O molecules.  The formation of
OH is a secondary step caused by photo-dissociation of H$_2$O by
internally produced UV photons.

Similarly, the high-density regions a) and b) do not harbor any 1612
MHz OH masers \citep{pihl2008}. This is expected since when the
density in the OH region increases towards values of 10$^{6}-10^{7}$
cm$^{-3}$, the 1720 MHz OH masers becomes quenched, and the 1612 MHz
transition instead becomes strongly inverted
\citep{lockett1999,pihl2008}. The lack of 1612 MHz masers in these
regions instead implies a chemistry where the OH abundance is
low. This may be explained by gas that is too dense for UV photons to
penetrate sufficiently deeply to create large amounts of OH from the
photo-dissociation of H$_2$O but keeping the methanol abundance
intact.

The above reasoning suggests that the relative distribution of maser
species is also a question of relative molecular abundance, rather
than simply density differences, emphasizing the need of observing
several different molecular transitions to probe the conditions across
these regions.

\begin{figure} 
  \includegraphics[scale=.43]{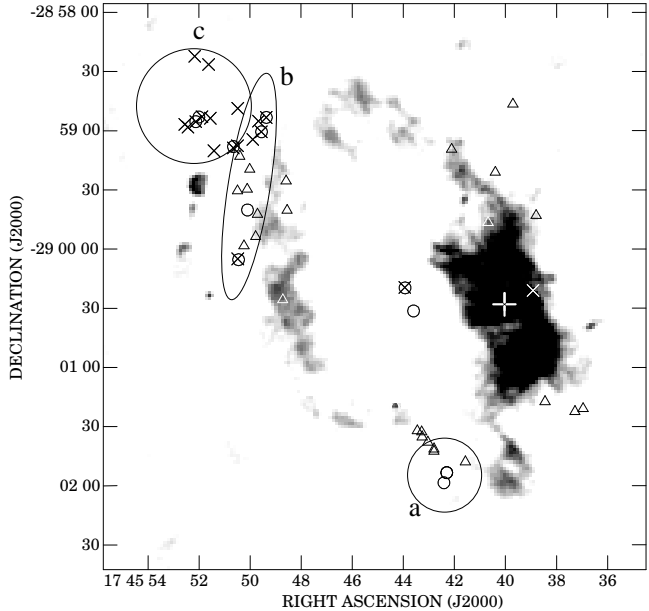}
\caption{Positions of 36 GHz CH$_3$OH masers (circles), 44 GHz
  CH$_3$OH masers (crosses), and 1720 MHz OH masers (triangles)
  overlayed on a 1.7~GHz continuum image of the Sgr\,A\,East
  region. The plus sign represents the position of Sgr A*. Only part
  of the SNR extent was searched for masers, for details see
  \citet{pihl2011}. Three different regions of methanol masers are
  defined: a) a couple of 36~GHz masers in the south, b) the linear group of co-spatial 36~GHz and 44~GHz masers in the north-east, and c) a group of 44~GHz dominated masers in the north-east corner.}
\end{figure}

\subsection{84~GHz and 95~GHz Methanol Masers in SNRs}

To date, 84~GHz CH$_3$OH maser emission has not been detected toward
any SNR, and there appears to be no published
systematic searches for this transition towards SNRs.
\citet{shulga2012} report a 95~GHz maser detection toward the SNR Kes
79, speculating that the maser emission is produced from the
interaction of the SNR and a neighboring molecular cloud. The 95~GHz
emission is brightest at densities around $10^5$ cm$^{-3}$, but
observations of additional transitions will be necessary in order to
further constrain the density and temperature. Kes~79 has been
detected in the 1720 MHz OH line, although not confirmed as maser
emission \citep{green1999}. The 95~GHz detection is a good example of
methanol being a complimentary shock tracer to the OH maser line.

\section{Summary}

Recent observations of methanol maser lines in SNRs have lacked a more
detailed understanding of the density regimes under which the
transitions occur. In this paper we presented level population
calculations of the 36, 44, 84, and 95 GHz methanol lines. These
calculations are used as a diagnostic tool to interpret detections and the
relative strength of methanol transitions in SNRs.

While the results can be used for probing the physical conditions of
individual regions where methanol masers are detected, they also provide a few summarizing results:

\begin{itemize}

\item Given a sufficient methanol abundance, methanol maser emission
  is obtained over a large range of densities and temperatures.
\item Line ratios are significantly changing with altering physical
  conditions, and regions where multiple transitions are observed allow firmer constraints on the physical parameters.
\item Detection of these methanol lines may serve as a complimentary
  shock tracer in SNRs to the commonly used 1720 MHz OH line. Compared
  to the OH masers, the velocities of the methanol masers are also more in alignment with
  the velocities of the cloud.
\item By using the calculated line ratios, observations of methanol
  masers in G1.4$-$0.1, W28, and Sgr\,A\,East illustrate that a wide
  range of physical conditions may exist between different
  SNR/molecular cloud interaction regions, and also within individual
  interaction regions.
  
\end{itemize}

Future work with observing 84~GHz and 95~GHz transitions towards
regions where 36~GHz and 44~GHz emission is observed may help
confirm these modeling results and conclusions.

\acknowledgments We would like to thank Andr\'es Ramos and Moshe
Elitzur for their help with MOLPOP-CEP. We thank NASA for support
under FERMI grant NNX12AO77G. B.C.M.\,also acknowledges support from
the NM Space Grant Consortium under the Graduate Research Fellowship program.
The National Radio Astronomy Observatory is a facility of the National
Science Foundation operated under cooperative agreement by Associated
Universities, Inc.

\end{document}